\documentclass[pra,showpacs,twocolumn]{revtex4}
\usepackage[english]{babel}
\usepackage{setspace}
\usepackage{amssymb}
\usepackage{amscd}
\usepackage{rotating}
\usepackage{color}
\usepackage{graphicx}
\begin{document}
\title{Boson Core Compressibility}
\author{Y. Khorramzadeh, Fei Lin, and V. W. Scarola}
\affiliation{Department of Physics, Virginia Tech, Blacksburg, Virginia 24061}
\begin{abstract}
Strongly interacting atoms trapped in optical lattices can be used to explore phase diagrams of Hubbard models.  Spatial inhomogeneity due to trapping typically obscures distinguishing observables.  We propose that measures using boson double occupancy avoid trapping effects to reveal two key correlation functions.  We define a boson core compressibility and core superfluid stiffness in terms of double occupancy.  We use quantum Monte Carlo on the Bose-Hubbard model to empirically show that these quantities intrinsically eliminate edge effects to reveal correlations near the trap center.  The boson core compressibility offers a generally applicable tool that can be used to experimentally map out phase transitions between compressible and incompressible states.  
\end{abstract}
\date{\today}

\maketitle

\section{Introduction}

Significant progress in cooling and trapping cold atomic gases in optical lattices \cite{verkerk:1992,jessen:1992,hemmerich:1993} established new and ideal platforms to study quantum condensed matter \cite{jaksch:1998,greiner:2002,bloch:2008}.  Ongoing work seeks to explore properties of interesting but  poorly understood quantum many-body states using quantum degenerate atoms.  Proposals include the use of optical lattice bosons to study novel superfluid order in higher bands \cite{scarola:2005,isacsson:2005,wu:2006} or topological phases \cite{duan:2003,micheli:2006,tewari:2006}.  Fermi gases in optical lattices are also under study as a route to explore the controversial phase diagram of the Fermi-Hubbard model \cite{hofstetter:2002,bloch:2008,esslinger:2010}.  

Time of flight observables in optical lattice experiments can be adapted to measure properties of atoms trapped in optical lattices.  The  momentum distribution \cite{greiner:2001}, density-density correlation functions \cite{folling:2005,spielman:2007}, compressibility \cite{jordens:2008,schneider:2008}, and double occupancy \cite{rom:2004,stoferle:2006,bakr:2010,sherson:2010} are all examples of working optical lattice observables.  At first it may seem that these observables can be used to directly pinpoint locations on phase diagrams because input parameters (e.g., lattice depth) are controlled and tunable.  But significant spatial inhomogeneity due to trapping can spoil the connection between phase diagrams and experiments.  

Recent experiments \cite{jordens:2008} with fermions ($^{40}\text{K}$) trapped in optical lattices used the double occupancy as an indicator of the Fermi-Gas to Mott insulator transition to circumvent issues due to trapping.  These experiments used a Feshbach resonance to shift hyperfine levels of doubly occupied sites.  Doubly occupied sites were promoted to a separate hyperfine state and measured in time of flight using a Stern-Gerlach scheme to distinguish atoms originating from singly and doubly occupied sites.  By taking the derivative of the double occupancy with respect to particle number these experiments effectively extracted the core compressibility of atoms in the optical lattice.  The core compressibility revealed that the center of the sample became incompressible as interactions tuned the system from a Fermi-Gas to a Mott insulator. 

Measurements of the optical lattice fermion double occupancy have been compared with theory.  A high temperature series expansion \cite{scarola:2009} was used to show that the fermion core compressibility measured in Ref.~\onlinecite{jordens:2008} does indeed capture the compressibility of the center of the system, even in the presence of severe spatial inhomogeneity.  Comparisons between high temperature series expansions, dynamical mean-field theory, and experiment were also useful in using double occupancy to measure the temperature of fermions in optical lattices \cite{jordens:2010}.  More recent calculations have shown that the fermion core compressibility can be very useful in one dimensional optical lattices as well \cite{hu:2010,snyder:2011}.

Here we explore the potential uses of double occupancy in measuring the compressibility of \emph{bosons}.  At first it may appear that boson double occupancy might not offer useful information (as it did for fermions) because several bosons can occupy a single site even in the same Bloch band.  We explore this issue through extensive calculations.  We find that, in the presence of strong boson-boson repulsion and at low temperature, double occupancy does offer a useful tool that can be related to important observables.  

We use the boson double occupancy to define the boson core compressibility and core stiffness. 
We find that these quantities intrinsically exclude edge effects thus offering valuable probes of a single phase near the system center.   We use quantum Monte Carlo (QMC) on the trapped Bose-Hubbard model to empirically show that the boson core compressibility and stiffness reveal the compressibility and stiffness of atoms near the trap center, respectively.   We show that the core compressibility defined here can be used in experiments to study critical properties and map out phase diagrams of any Bose-Hubbard model even in the presence of significant spatial inhomogeneity due to trapping.  The core compressibility can be used, for example, to explore transitions in disordered Bose-Hubbard models currently under experimental investigation \cite{fallani:2007,white:2009,pasienski:2010}.  Our proposed boson core compressibility complements proposals to measure total boson compressibility with trap squeezing \cite{delande:2009,roscilde:2009}.

In Section~\ref{model} we discuss the Bose-Hubbard model as a relevant testbed for computing the boson core compressibility.  In Section~\ref{local} we define the local compressibility and the local superfluid stiffness.  In Section~\ref{seciontdefinecore} we define the core compressibility and the core stiffness.  In Section~\ref{uniform} we use mean-field theory (MFT) and QMC to compare the global quantities with core quantities in uniform systems.  Section~\ref{qmctrap} uses QMC on trapped systems to compare the local compressibility and the local stiffness with the core compressibility and the core stiffness, respectively.  We find parameter regimes where core measures essentially track local quantities near the trap center.  Section~\ref{measuring} discusses how to effectively extract the boson core compressibility from time of flight measurements.

\section{Model}
\label{model}

\begin{figure}[t]
  \includegraphics[width=6.0cm, height=4.8cm]{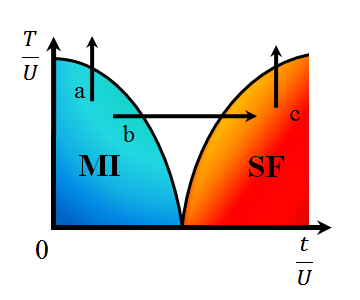}
  \caption{(Color online) Schematic of the finite-temperature phase diagram of the uniform Bose-Hubbard model. The Mott insulator (MI) and the superfluid (SF) are separated by a narrow quantum critical regime at finite temperature.  At high temperatures the system is in the normal phase.  The lines represent: a) a transition from the Mott insulator to the normal phase, b) a transition from the Mott insulator, through the quantum critical regime, into the superfluid, and c) a transition out of the superfluid into the normal phase. }
  \label{phasediagram}
\end{figure}

The Bose-Hubbard models offers one of the simplest models with a quantum phase transition \cite{fisher:1989,sachdev:1999}.  It also captures the essential properties of many ongoing optical lattice experiments \cite{jaksch:1998,bloch:2008}.  We may therefore use the Bose-Hubbard model as an experimentally relevant testbed to examine the usefulness of the boson core compressibility:  
\begin{equation}
 H = -t\sum_{\left \langle i,j \right \rangle} (b_{i}^{\dagger} b_{j}^{\phantom\dagger}+  H.c.)+\frac{U}{2}\sum_{i}n_{i}\left ( n_{i}-1 \right ) - \sum_{i} \mu_{i} n_{i}.
 \label{bhmodel}
\end{equation}
Here $n_{i}= b_{i}^{\dagger} b_{i}^{\phantom\dagger}$ is the number operator at a lattice site indexed by $i$ and $\mu _{i}=\mu -\gamma R_{i,0}^{2}$ is the local chemical potential.  The central chemical potential, $\mu$, tunes the average density, $\gamma$ parameterizes the parabolic confinement potential, and $R_{i,0}\equiv \vert {\bf R}_{i} - {\bf R}_{0}\vert $ is the distance between a site at ${\bf R}_{i}$ and the center of the trap, ${\bf R}_{0}$ . We work in units such that $U = 1$ and the lattice spacing is also set to unity.  In the following we will work on a simple cubic lattice with periodic boundaries.

In the uniform limit ($\gamma=0$) the model exhibits a quantum phase transition between a Mott insulator and a superfluid \cite{fisher:1989}.  Fig.~\ref{phasediagram} shows a schematic of the finite temperature  phase diagram at fixed $\mu$.  The Mott insulator is characterized by an integer density enforced by the Mott energy gap $\sim U$.  As a result of the energy gap, the Mott insulator is incompressible.  The superfluid phase is compressible.  It is characterized by strong number fluctuations (even at low temperature) and a finite superfluid density.  

\begin{figure}[t]
  \includegraphics[width=6.0cm, height=4.8cm]{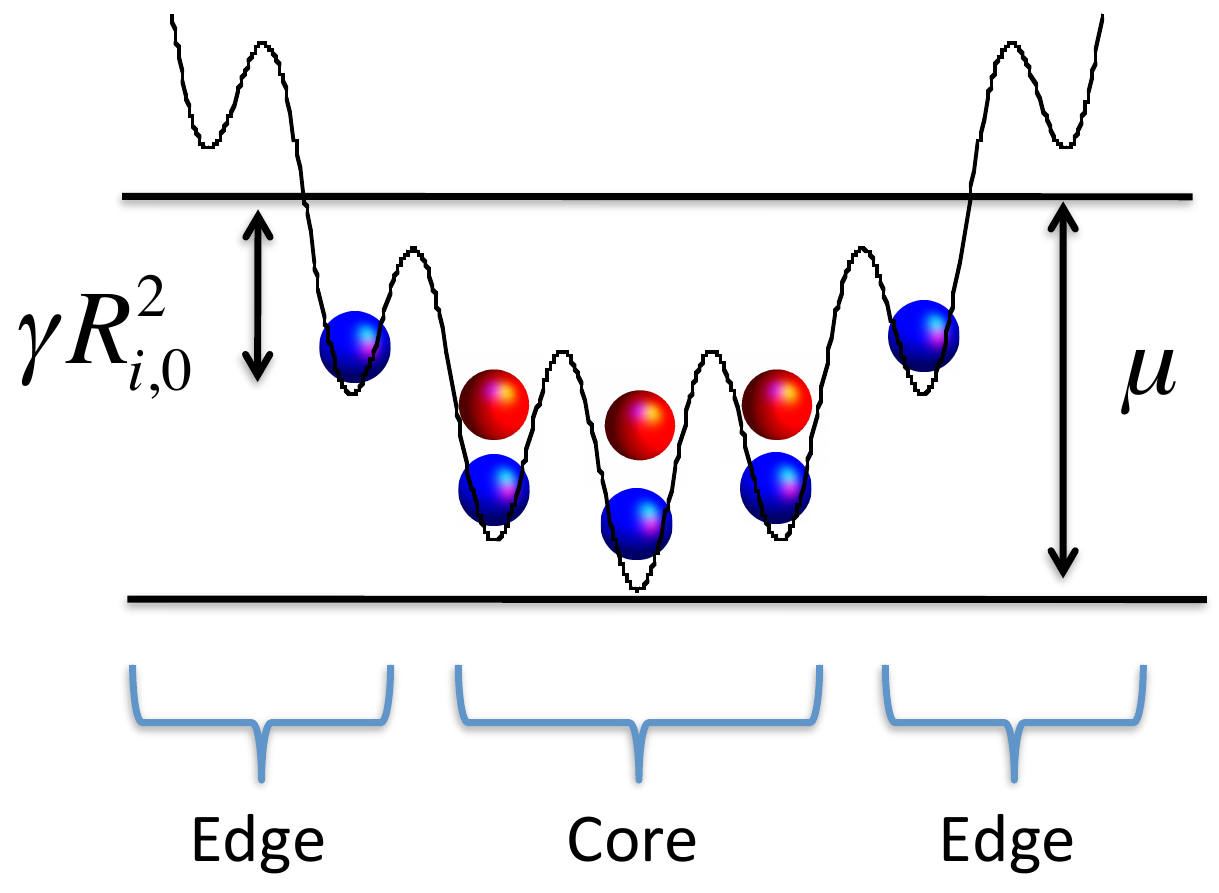}
  \caption{(Color online) Schematic depicting spatial inhomogeneity due to trapping in an optical lattice.  A spatially varying chemical potential ($\gamma R_{i,0}^{2}$) modifies the otherwise uniform chemical potential, $\mu$ , to model a parabolic confinement potential.  The center of the trapped system is denser than the edges.  In the absence of large thermal and quantum fluctuations, double occupancies tend to cluster around the system center.  The core compressibility implicitly defines the core region as the region with finite double occupancy in the system.  }
  \label{schematic_trap}
\end{figure}

Trapping due to confinement magnetic fields and/or tapered laser beam waists in optical lattice experiments are modeled by a parabolic trapping potential (Fig.~\ref{schematic_trap}). The last term in Eq.~(\ref{bhmodel}) decreases the density as the distance from the center of the system increases for $\gamma >0$.  As a result the spatially varying chemical potential mixes phases within the trap.  Identifying individual phases within the trapped system requires a local observable but most experiments currently rely on bulk time of flight imaging.  We study the boson core compressibility as a candidate quasi-local observable that uses bulk time of flight data to measure a single phase within the trap.  Core quantities can be compared with single-site local quantities to show that only one phase in the trap is measured.

\section{Local Compressibility and Local Superfluid Stiffness}
\label{local}

Time of flight observables relate to correlation functions, e.g., momentum distribution, typically computed in uniform systems.  In this section we first define bulk compressibility and bulk superfluid stiffness.  We then adapt these definitions to specific measures of local compressibility and local superfluid stiffness.

According to fluctuation-dissipation theorem, density fluctuations in an optical lattice contain useful information about the system 
\cite{duchon:2011, zhou:2011, gemelke:2009, ma:2010, fang:2011}.
The total compressibility, $\kappa$, measures the ability of the system to change its density with small changes in the chemical potential:\begin{equation}
 \kappa=\frac{\partial \left \langle n \right \rangle}{\partial \mu}=N_s^{2}\beta \left [ \left \langle n^{2} \right \rangle -\left \langle n \right \rangle^{2} \right ],
\label{kappatotal}
\end{equation}
where $\left \langle n^{2} \right \rangle=N_s^{-2} \langle  (\sum_{i=1}^{N_s} n_{i})^{2}\rangle$, $\left \langle n \right \rangle=N_s^{-1}  \langle \sum_{i=1}^{N_s}n_{i} \rangle$ and $\beta = (k_{B}T)^{-1}$ denotes the inverse temperature. We set $k_{B}=1$ in the following.  $ N_s=L^{3}$ is the total number of sites. Angular brackets denote the thermal average of observables: $<A> =\text{Tr}\left \{A e^{-\beta H } \right \}/Z$, where $Z$ is the grand canonical partition function, $Z=\text{Tr}\left \{ e^{-\beta H } \right \}$, and $\text{Tr}$ denotes the trace.  The last equality in Eq.~(\ref{kappatotal}) shows that the compressibility is intrinsically non-local because it relates to density fluctuations across the entire system. 

Bose-Einstein condensation in the presence of interactions leads to superfluidity \cite{pines:1966}.  The superfluid stiffness (and therefore the superfluid density) can be computed using the response of the system to weak perturbations and, in turn, the winding number, $W$, evaluated across the system's boundary in a QMC simulation \cite{pollock:1987}:
\begin{eqnarray}
 \rho_{s}=\frac{W^{2}}{2t\beta\langle n \rangle}.
 \end{eqnarray}
The superfluid stiffness, as defined, is a manifestly bulk quantity because macroscopic occupation of a single mode in the presence of interactions implies that perturbations lead to non-local response.

Significant spatial inhomogeneity due to trapping suggests that \emph{local} measures of compressibility and stiffness will be more informative.  The local compressibility \cite{wessel:2004}:
\begin{equation}
 \kappa_{i}=\frac{\partial \left \langle n \right \rangle}{\partial \mu _{i}}=\beta \left[ \left \langle n_{i}n \right \rangle -\left \langle n_{i} \right \rangle\left \langle n \right \rangle\right],
 \label{localk}
\end{equation}
measures the average density fluctuations at a single site in comparison to the total average density.  Note that the local compressibility of a trapped system becomes equivalent to $\kappa$ in a uniform system only if $\kappa_{i}$ is summed over a smooth and large volume of the trapped system.  For large enough system sizes the local compressibility shows critical properties, similar to those of the total compressibility, that can be used to identify phase boundaries \cite{rigol:2003,wessel:2004}.

Figure~\ref{schematic_trap} depicts severe inhomogeneity imposed by a trapping potential.  Here we see that the compressibility in the core of the system can be entirely different from the edges.  We thus expect $\kappa\neq \kappa_{i_{\text{c}}}$, where $i_{c}$ denotes a site at the trap center.

To capture local order of trapped bosons it is also convenient to define a local stiffness.  We define the local stiffness in terms of a projection along imaginary time in QMC simulations.  We calculate the superfluid stiffness using the system's response to weak rotation \cite{ceperley:1995} measured in imaginary time. Since particles in the superfluid state do not respond to rotation, only particles in the normal state contribute to the system's total moment of inertia, $I$. The superfluid stiffness is then equivalent to the ratio of the missing moment of inertia to the classical 
moment of inertia $I_{\text{cl}}$ of the system, i.e., 
\begin{equation}
\rho_s=\frac{I_{\text{cl}}-I}{I_{\text{cl}}}.
\end{equation}
This relation offers a physical interpretation of superfluid stiffness that can be used to construct a local stiffness.

We use the QMC formalism to construct the local superfluid stiffness. In the QMC formalism the total stiffness can be expressed as: 
 \begin{equation}
 \rho_s^{\alpha}=\frac{4\mathcal{M}^2\langle A_{\alpha}^2\rangle}{\beta\hbar^2I^{\alpha}_{\text{cl}}},
 \label{totalstiffness}
 \end{equation}
 where $\alpha=x,y,z$ is the rotation axis through the system center, 
\begin{equation}
I^{\alpha}_{\text{cl}}=\left \langle \mathcal{M}\sum_{l,\tau=1}^{N,\tau^{m}}{\bf R}_{\perp}(l,\tau)\times {\bf R}_{\perp}(l,\tau+1)\right \rangle,
\end{equation}
and
\begin{eqnarray}
 A_{\alpha} ({\bf R}_{i})=\frac{1}{2}\sum_{l,\tau=1}^{N,\tau^{m}}\left[{\bf R}(l,\tau)\times {\bf R}(l,\tau+1)\right]_{\alpha}\delta_{{\bf R}_{i},{\bf R}(l,\tau)}\nonumber
 \end{eqnarray}
is the projected area when $N$ particles of mass $\mathcal{M}$ move along paths in the imaginary-time direction in QMC. Here  ${\bf R}(l,\tau)$ 
denotes the site of the $l^{\text{th}}$ particle at the $\tau^{\text{th}}$ imaginary-time step while ${\bf R}_{\perp}(l,\tau)$ denotes the same but for the distance between the particle and the principal axis, $\alpha$. $\tau^{m}$ is the maximum number of imaginary time steps and $\delta$ denotes the Kronecker delta.  The total projected area becomes:
\begin{equation}
A_{\alpha}=\frac{1}{N_{s}}\sum_{i}A_{\alpha} ({\bf R}_{i}).
\end{equation}
Note that in above we have defined a particle mass
$\mathcal{M}=\hbar^2/2ta^2$ and lattice spacing $a=1$.

We use the above local quantities to define a local superfluid stiffness.  The definition of local superfluid stiffness is not unique. 
Here we use the definition of Ref.~\onlinecite{kwon:2006}:
\begin{equation}
 \rho_s({\bf R}_{i})=\frac{2\langle A_{\alpha}A_{\alpha}({\bf R}_{i})\rangle}{t\beta {{\bf R}_{i}}^2_{\perp}},
 \label{localstiffness}
\end{equation}
which, if multiplied by $\mathcal{M} {{\bf R}_{i}}^2_{\perp}$ and integrated over all the lattice sites, yields the total 
superfluid stiffness of the system, Eq.~(\ref{totalstiffness}).  Eq.~(\ref{localstiffness}) allows us to pinpoint the presence (or absence) of local superfluids in our simulations of trapped systems.

\section{Core Compressibility and Core Stiffness}
\label{seciontdefinecore}

Measurements of the double occupancy can be used to observe local quantities.  Due to confinement we expect the local density to be largest near the center of the trapped system.  This implies that, in the absence of significant quantum fluctuations, double occupancies should cluster near the core of the system.  In this section we first define decomposition into occupancies.  We then expand the density in terms of occupancy.  We use the total double occupancy to define the boson core compressibility and core stiffness.

Any operator $\mathcal{O}$ can be decomposed into a sum over occupancy in Fock space.  If we assume that $\mathcal{O}$ is a sum over local operators at a site, $\mathcal{O}_{i}$, we can define our decomposition in terms of projection operators:
\begin{eqnarray}
\mathcal{O}= \sum_{i=1}^{N_{s}}\sum_{m=0}^{\infty} \mathcal{O}_{i}  \mathcal{P}_{i,m},
\end{eqnarray}
where $\mathcal{P}_{i,m} $ projects onto the Fock states of the $i^{\text{th}}$ site:
\begin{eqnarray}
 \mathcal{P}_{i,m} \equiv \left( b_{i}^{\dagger} \right)^{m}\vert 0 \rangle \langle0 \vert \left ( b_{i}^{\phantom\dagger} \right)^{m}.
\end{eqnarray}
We can therefore rewrite the expectation value of an observable as a sum over the occupancy projectors:
\begin{eqnarray}
\langle \mathcal{O} \rangle = \left\langle \sum_{i,m} \mathcal{O}_{i}  \mathcal{P}_{i,m} \right\rangle.
\label{expansion}
\end{eqnarray}
For example, the total density is given by:
\begin{eqnarray}
\langle n \rangle =N_{s}^{-1} \left\langle \sum_{i,m} \left( b_{i}^{\dagger} \right)^{m} \vert 0 \rangle m \langle 0\vert \left ( b_{i}^{\phantom\dagger} \right)^{m}\right\rangle.
\label{densityexpand}
\end{eqnarray}
Using Eq.~(\ref{densityexpand}) we can express the total compressibility, $\kappa=\partial \langle n\rangle /\partial \mu$, as an expansion over occupancy.    
 
In trapped optical lattice experiments there is an approximate correspondence between occupancy and location within the trap.  Furthermore, occupancies can be readily measured (See, e.g., Refs.~\onlinecite{rom:2004,stoferle:2006,jordens:2008,bakr:2010,sherson:2010}, and~\onlinecite{jordens:2010}). The double occupancy arises from the $m=2$ term in Eq.~(\ref{densityexpand}): $b^{\dagger}_{i}b^{\dagger}_{i} b^{\phantom\dagger}_{i} b^{\phantom\dagger}_{i}=n_{i}(n_{i}-1)$. The total boson double occupancy is given by:
\begin{equation}
\langle D \rangle\equiv\frac{1}{2}\left \langle \sum_{i}^{N_s} n_{i}\left ( n_{i}-1 \right )\right \rangle.  
\end{equation}
The double occupancy per lattice site is $\langle d \rangle \equiv \langle D\rangle /N_{s}$.  Here we see that $ \left \langle n_{i}\left ( n_{i}-1 \right )\right \rangle$ is non-zero if there are at least two particles at a site.  Measuring double occupancies offers a global observable that yields local information by excluding edge effects (provided the edges have a low density).  Thus observables based on double occupancy measure properties at the core of trapped optical lattices.  

We define the double occupancy core compressibility by expanding $\partial \langle n\rangle /\partial \mu$ over the occupancies.  We find:
\begin{equation}
 \kappa_{c}^{d}\equiv \frac{1}{2N_s}\frac{\partial }{\partial \mu }
\left \langle \sum_{i}^{N_s} n_{i}\left ( n_{i}-1 \right )\right \rangle,
\label{corek}
\end{equation}
in direct analogy to a similar measure used for fermions \cite{jordens:2008,scarola:2009}.  Here small changes in chemical potential impact $d$ only if the doubly occupied sites form a compressible state.  We will show, by direct calculation, that $ \kappa_{c}^{d}$ offers a quantitatively accurate estimate of the compressibility near the center of a trapped optical lattice system of bosons.  

Eq.~(\ref{corek}) can be generalized to measure the compressibility near a state of any density.  The $m=3$ term in Eq.~(\ref{densityexpand}) gives the triple occupancy term in the density expansion. A measurement of triple occupancy can be used to observe the compressibility near the center of a trapped system with triply occupied sites near the trap center:
\begin{equation}
 \kappa_{c}^{t}\equiv  \frac{1}{6N_s} \frac{\partial } {\partial \mu } \left \langle \sum_{i}^{N_s}n_{i}\left ( n_{i}-1 \right )\left( n_{i}-2 \right )\right \rangle.
\end{equation}
$\kappa_{c}^{t}$ excludes both singly and doubly occupied sites.

The expansion of observables in terms of occupancies (Eq.~\ref{expansion}) is a general procedure that can be applied to other order parameters.  We also define a core superfluid stiffness in terms of doubly occupied sites.  We modify Eq.~(\ref{totalstiffness}) to incorporate only doubly occupied sites by redefining the location of a particle in QMC, ${\bf R}(l,\tau)\rightarrow{\bf R}^{d}(l,\tau)$, where ${\bf R}^{d}(l,\tau)$ defines the site of the $l^{\text{th}}$ particle in imaginary time provided it sits on a doubly occupied site.  The global stiffness then reduces to the superfluid density of doubly occupied sites.
We will see that, in a trap, this measures the superfluid density near the core of the sample.  The core superfluid density, $\rho^{d}_{s}$, can be compared with the local superfluid density,  $\rho_s({\bf R}_{i})$.  We will use QMC to show quantitative agreement between both quantities in the core of the Bose-Hubbard model.

We expect $\kappa_{c}^{d}$  and $\rho^{d}_{s}$ to give an accurate measure of local compressibility and local stiffness near the trap center in a specific but interesting regime of any Bose-Hubbard model:  i) The density should be just above unity, ii) The temperature should be low enough to prevent a significant number of triply occupied sites near the center or doubly occupied sites near the edges, and iii) Quantum fluctuations should not be strong enough to induce triply occupied sites near the center or doubly occupied sites near the edges.  We will use MFT and QMC on the Bose-Hubbard model to demonstrate that these criteria can indeed be satisfied.

\section{Uniform Systems}
\label{uniform}

Intuition and understanding of quantum many-body systems often play out in large-uniform systems where translational invariance simplifies assumptions.  The connection between optical lattice observables and conventional order parameters defined in uniform systems can be tenuous because of trapping effects.  In this section we establish a quantitative connection between core compressibility and total compressibility in uniform systems.  We will show that the Bose-Hubbard model offers several regimes where the core compressibility can be used to observe total compressibility.  In Section~\ref{qmctrap} we then turn to comparisons between the core and local compressibility in realistic trapped systems.

We begin our comparison between core compressibility and total compressibility in a regime where MFT applies.  Consider the line marked ``a'' in Fig.~\ref{phasediagram}.  For weak hopping $t$ (or in the absence of superfluidity) it is sufficient to ignore the hopping term in the Hamiltonian.  This approximation is not as severe as it appears.  At the mean-field level our weak hopping approximation follows from two steps.  i) We first decouple sites \cite{sachdev:1999}:
\begin{eqnarray}
-t\sum_{\left \langle i,j \right \rangle} (b_{i}^{\dagger} b_{j}^{\phantom\dagger}+  H.c.)\rightarrow -t\sum_{i} \left( \psi^{*} b_{i}^{\phantom\dagger} + \psi b_{i}^{\dagger} \right )
\end{eqnarray}  
where $\psi\equiv\langle b \rangle$ is the mean-field superfluid order parameter.  ii) In the absence of a finite superfluid density we can take $\psi=\psi^{*}=0$.  In this case the mean-field hopping term vanishes and we can focus on the remaining diagonal terms in the Bose-Hubbard model.

In the absence of mean-field superfluidity, the mean-field Hamiltonian becomes: 
\begin{eqnarray}
H^{MF} =\frac{U}{2} \sum_{i}n_{i}  \left ( n_{i} -1 \right )- \sum_{i}\mu_{i}  n_{i} 
\label{meanfieldmodel}
\end{eqnarray}
In this limit the model is diagonal and can be solved exactly.  The total mean-field energy of the $\psi=0$ system is then: 
$$E = \sum_{i=1}^{N_s} \left [\frac{U}{2}\left \langle n_{i} \right \rangle\left (\left \langle n_{i} \right \rangle-1  \right )-\mu _{i}\left \langle n_{i} \right \rangle \right ]$$
Note that the energy increases with $\langle n_{i} \rangle$.  We can count energy eigenvalues using integer local number occupancy: $E_m=Um(m-1)/2-\mu m$, with $m=0,1,2,...$.   In regimes where we can separate energy scales according to occupancy our definition of core compressibility becomes an accurate representation of the total compressibility because we can locate parameter regimes where singly and triply occupied sites can be ignored.  

The mean-field limit described above provides a limit in which the core compressibility exactly captures the compressibility of doubly occupied sites.  To construct a comparison between total and core compressibility measures note that in the mean-field limit we rewrite Eq.~(\ref{densityexpand}):
\begin{equation}
\langle n \rangle \approx \sum_{m=0}^{\infty}m \delta \rho_{m}
\label{mftdensity}
\end{equation}
where the thermodynamic factor $\delta\rho_{m} \equiv \exp \left\{ -\beta E_{m}\right \}/Z$ gives a deviation in density from an integer value at finite temperatures.  Eq.~(\ref{mftdensity}) becomes exact for $t=0$.  

Using density deviations we can relate the compressibilities defined in Sec.~\ref{seciontdefinecore}.  The density, double occupancy, and triple occupancy can all be decomposed in terms of density deviations factors:
\begin{eqnarray}
\langle n \rangle &\approx& \delta \rho_{1}+2\delta \rho_{2}+3\delta \rho_{3}+... \nonumber \\
\langle n(n-1)\rangle/2 &\approx& 0+\delta \rho_{2}+3\delta \rho_{3}+...\nonumber \\
\langle n(n-1)(n-2)\rangle/6 &\approx& 0+0+\delta \rho_{3}+... \nonumber
\end{eqnarray}
By taking derivatives with respect to the total chemical potential we arrive at three types of compressibilities:
\begin{eqnarray}
 \kappa \ &\approx& \frac{\partial \delta \rho_{1}}{\partial \mu}+2\frac{\partial \delta \rho_{2}}{\partial \mu}+3 \frac{\partial \delta \rho_{3}}{\partial \mu}+...\nonumber \\
 \kappa_{c}^{d}  &\approx& 0+\frac{\partial \delta \rho_{2}}{\partial \mu}+3\frac{\partial \delta \rho_{3}}{\partial \mu}+... \nonumber\\
\kappa_{c}^{t} &\approx& 0 + 0+ \frac{\partial \delta \rho_{3}}{\partial \mu}+...\nonumber
\end{eqnarray}
With these expansions we see explicitly that the core compressibilities give the compressibility of doubly and triply occupied sites while ignoring sites with lower occupancy.  

Using the density deviation expansion, we can find regimes for which $\kappa\approx\kappa_{c}^{d}$.  We require parameters such that:
$$  \frac{\partial \delta \rho_{1}}{\partial \mu}\approx - \frac{\partial \delta \rho_{2}}{\partial \mu} 
\text{and} \left \vert \frac{\partial \delta \rho_{3}}{\partial \mu}\right \vert  \ll \left\vert \frac{\partial \delta \rho_{2}}{\partial \mu}\right \vert. $$
The first requirement holds for chemical potentials just large enough to add a small number of vacancies to the Mott insulator.  For example, consider the  limit $\langle n\rangle =1+\epsilon$, where $\epsilon\ll1$.  The first condition holds because vacancies have nearly equal and opposite compressibility as doubly occupied sites.  The second equality holds for temperatures and chemical potentials low enough to prevent significant triple occupancy.  Thus an experiment measuring double occupancy can approximate the mathematical procedure of projection into occupancies defined in Eq.~(\ref{densityexpand}).

We have shown that the core compressibility defines a compressibility of doubly occupied sites at the mean-field level.  We have also argued that regimes near the Mott insulator yield $\kappa\approx\kappa_{c}^{d}$.  The arguments made using the mean-field approximation are exact at $t=0$ (or with $\psi=0$ in the absence of quantum fluctuations). 

To find regimes where $\kappa\approx\kappa_{c}^{d}$ in other parts of the phase diagram we compute both quantities using QMC.  Correlation functions computed with QMC on the Bose-Hubbard model are numerically exact on finite sized systems.  We use the stochastic series expansion representation with directed loop updates \cite{syljuasen:2002} to evaluate observables of Eq.~(\ref{bhmodel}).  We work within the ALPS framework \cite{albuquerque:2008} for some of the calculations.  Error bars for all plotted data points are smaller than the size of the symbols used within the figures unless otherwise indicated.

\begin{figure}[t]
  \centering
  \includegraphics[width=0.5\textwidth]{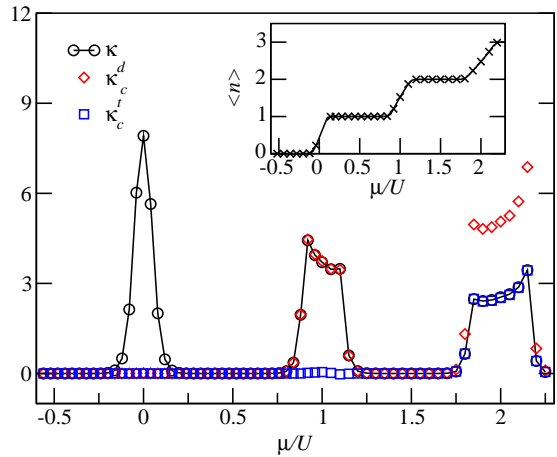}
  \caption{(Color online) The symbols show quantum Monte Carlo results for three different compressibilities computed in the uniform Bose-Hubbard model with $T/U=0.025$, $t/U=0.01$, and $L=6$ as a function of the chemical potential.  The total compressibility (circles) vanishes in the Mott insulator state but jumps when the chemical potential passes through the superfluid state.  The double occupancy-based core compressibility (diamonds) match the total compressibility when the density is around 1 and below 2. It vanishes for density less than 1.  The triple occupancy-based core compressibility (squares) match the total compressibility when the density is around 2 and below 3. It vanishes for density below 2.  Inset:  The total density plotted versus chemical potential for the same parameters as the main panel.  The solid lines in both the main panel and the inset are guides to the eye.}
  \label{uniformvsmu_withMFTresults}
\end{figure}

Fig.~\ref{uniformvsmu_withMFTresults} shows QMC results comparing the total compressibility (circles with solid line) with the core compressibilities as a function of chemical potential.  The inset shows that the chemical potential sweeps through Mott insulators at $\langle n \rangle=0,1,2$ and $3$.  Between Mott insulators the system forms compressible superfluids (peaks in the main panels).  The diamonds show that $\kappa_{c}^{d}$ tracks $\kappa$ \emph{only} for densities between 1 and 2.  Otherwise $\kappa_{c}^{d}\approx0$ when the system shows few triple occupancies ($\mu/U\lesssim1.75$).   Above $\mu/U\approx1.75$ the core compressibility based on double occupancies fails to track the total compressibility.  Here we find instead,  $\kappa\approx\kappa_{c}^{t}$.  Fig.~\ref{uniformvsmu_withMFTresults}  reveals two key features of  $\kappa_{c}^{d}$: i) there is a range of chemical potentials for which we find $\kappa\approx\kappa_{c}^{d}$ and ii) for all lower chemical potentials we find $\kappa_{c}^{d}\approx0$.  Thus edge effects induced by trapping (which lowers the chemical potentials near the edges) will have very little impact on $\kappa_{c}^{d}$ in comparison to $\kappa$.  The vanishing of the core compressibilities at low $\mu$ is a key property that allows the exclusion of edge effects in trapped systems when measuring $d$ for the entire trapped system.

\begin{figure}
  \centering
  \includegraphics[width=0.5\textwidth]{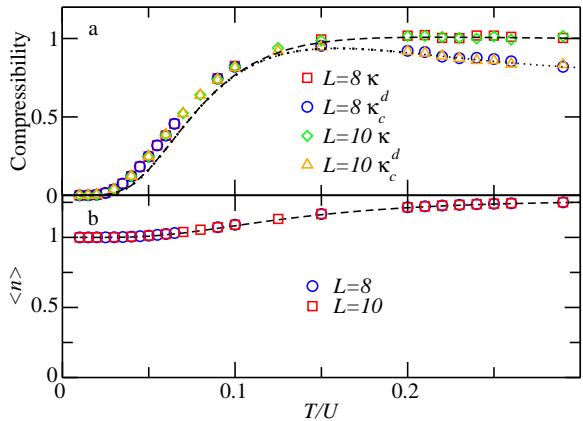}
  \caption{(Color online) Panels (a) and (b) compare quantum Monte Carlo (symbols) with mean-field results (lines), versus temperature for the compressibility and average density, respectively for the uniform Bose-Hubbard model.  Parameters are chosen in a regime where mean-field theory agrees with quantum Monte Carlo: at small hopping $ t = 0.01$ and $\mu=0.76$.  In panel (a) the dashed line is computed for the total compressibility but the dotted line is core compressibility.  This mean-field regime emphasizes the close agreement between the core and total compressibilities at low temperatures ($T/U\lesssim 0.15$).  There is even close agreement as we cross from the Mott insulator to the normal phase.  At higher temperatures ($T/U\gtrsim 0.15$), thermal fluctuations induce triple occupancies which causes the core and total compressibilities to deviate.  In panel (b) we see that the average density deviates from unity when temperatures are high enough to convert the Mott insulator into the normal phase.  
}
  \label{uniformvsTemperature_nospace}
\end{figure}

Fig.~\ref{uniformvsTemperature_nospace}  compares MFT with QMC for the density, the compressibility, and the core compressibility in a uniform system.  Parameters were chosen to host an incompressible Mott insulator at low temperatures.  Here we move along line ``a'' in Fig.~\ref{phasediagram}.  Above $T/U\approx 0.025$ ($T/U\approx0.15$) thermal fluctuations induce double (triple) occupancies.   The bottom panel shows that the density starts to deviate from unity when the temperature is increased.  But the top panel shows that for $T/U\lesssim 0.15$ we find  $\kappa\approx\kappa_{c}^{d}$.  At larger temperatures significant triple occupancies spoil the connection between $\kappa_{c}^{d}$ and total compressibility.  The solid lines draw the MFT.  Here the MFT is nearly indistinguishable from the QMC.  We conclude that weak thermal fluctuations allow measurements of $\kappa$ with $\kappa_{c}$.

\begin{figure}[t]
  \centering
  \includegraphics[width=0.5\textwidth]{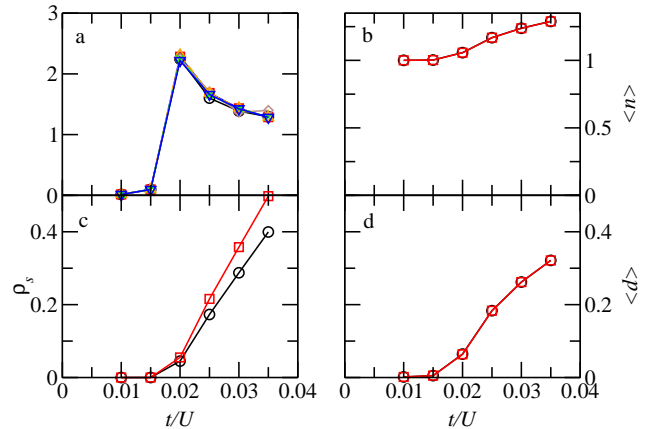}
  \caption{(Color online) Plots of thermodynamic quantities computed with quantum Monte Carlo as function of hopping for the uniform Bose-Hubbard model with $\mu=0.76$, $T=0.025$, and $L=8$ and $10$.  In panel (a), three different compressibilities are plotted: core compressibility ($\kappa_{c}^{d}$, squares for $L=8$ and up triangles for $L=10$), local compressibility ($\kappa_{i_{\text{c}}}$, circles for $L=8$ and diamonds for $L=10$), and total compressibility, ($\kappa$, down triangles for $L=8$ and X for $L=10$). In panels (b)-(d) circles denote $L=8$ data and squares $L=10$.  The solid lines are guides to the eye.  Panel (a) shows quantitative agreement between all measures of compressibility as we increase the hopping to drive the Mott insulator into the quantum critical regime.
}
  \label{uniformvsHopping}
\end{figure}

We also find that the core compressibility tracks the total compressibility even when there is not a clear connection between the density and energy, i.e., beyond the mean-field limit defined by Eq.~(\ref{meanfieldmodel}).  We choose parameters to move along the line ``b'' in Fig.~\ref{phasediagram}.  Here we start in a Mott insulator and move into the critical regime and then the superfluid phase where the stiffness is finite.  Fig.~\ref{uniformvsHopping} plots the stiffness, density, and double occupancy along with the compressibilities to show that the core compressibility still tracks the total compressibility even as the system experiences quantum fluctuations.  

\begin{figure}[t]
  \centering
  \includegraphics[width=0.5\textwidth]{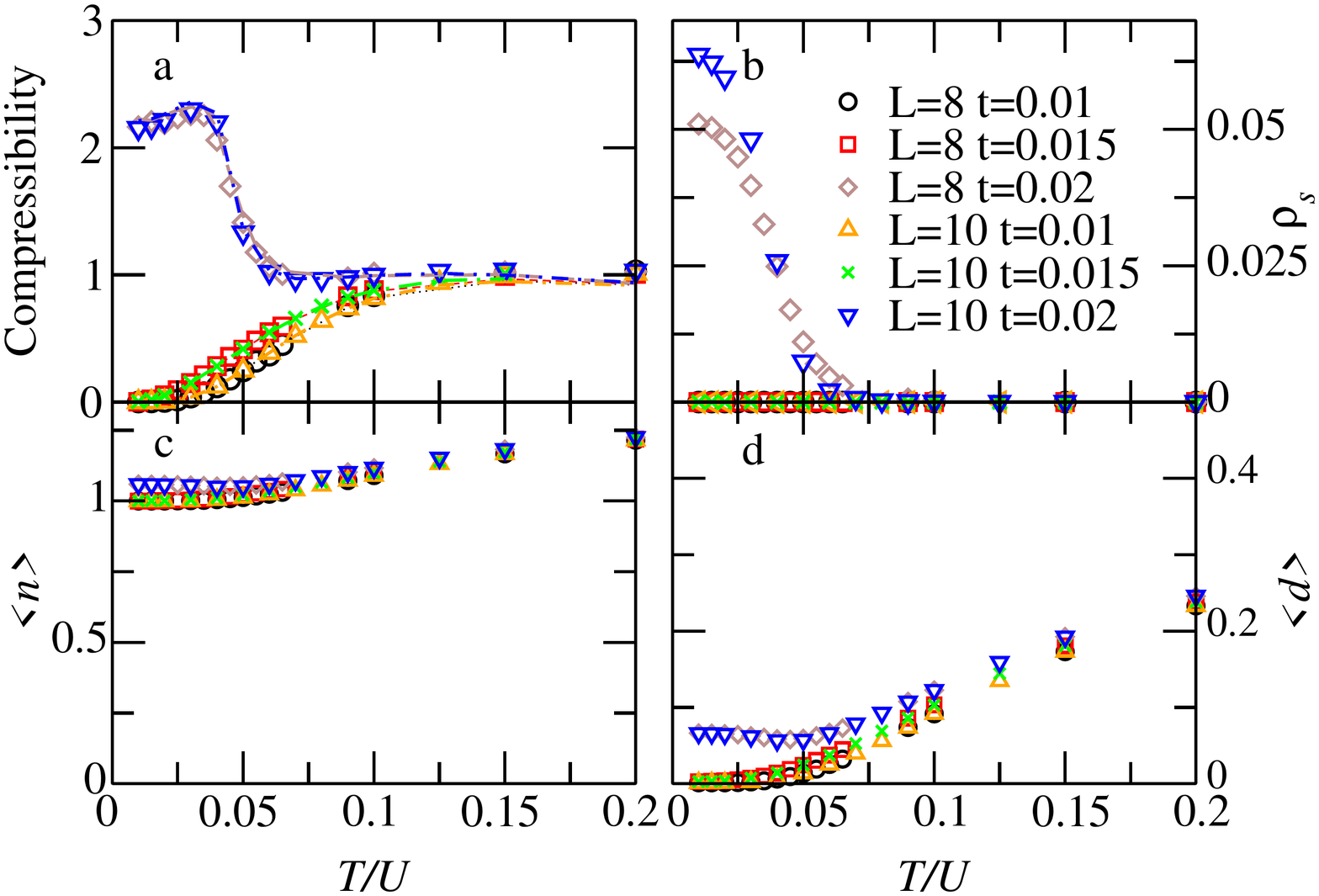}
  \caption{(Color online) Panels (a)-(d) plot quantum Monte Carlo results for the compressibility, superfluid stiffness, total density, and double occupancy, respectively, as a function of temperature for different lattice sizes and hoppings in the uniform Bose-Hubbard model. The chemical potential is fixed at $\mu/U=0.76$.  In panel (a) the total compressibility (symbols) shows agreement with the core compressibility, $\kappa_{c}^{d}$, (lines).  For low $t$ we see a transition from the Mott to the normal phase.  For large $t$ we see a transition from the superfluid to the normal phase.  The core compressibility reveals both transitions. }
  \label{uniformvsTemperature}
\end{figure}

We find similar agreement between $\kappa$ and $ \kappa_{c}^{d}$ deep in the superfluid regime.  There is no Mott insulator as we move along the line ``c'' in Fig.~\ref{phasediagram}.  Fig.~\ref{uniformvsTemperature} moves along both lines ``a'' and ``c'' in  Fig.~\ref{phasediagram}.  The compressibilities here also show agreement amongst themselves.  

In this section we used a MFT to argue that the core compressibility measures the compressibility near integer fillings in uniform systems.  We compared mean-field calculations with QMC to show that the core compressibility can be used to track phase transitions out of the incompressible Mott insulator state.  We also find that the core  compressibility agrees with the total compressibility even in the superfluid state provided we work at low densities.

\section{Trapped Systems}
\label{qmctrap}

We now turn to studies of the compressibility and stiffness in a  more realistic  setting, the trapped Bose-Hubbard model.   We consider regimes with non-zero $\gamma$ in Eq.~(\ref{bhmodel}).   We first show that 
core quantities accurately capture local quantities.  Specifically, we find regimes where $\rho^{d}_{s}$ and $\kappa_{c}^{d}$ accurately capture the local stiffness and the local compressibility, respectively, in the center of the trap.  We then construct a recipe for scaling the core compressibility.  By increasing particle number experiments can be used to access the thermodynamic limit if measurement data can be appropriately scaled with system size.  We show that the core compressibility can be scaled with trap strength.  We conclude that the core compressibility can be used to extract the total compressibility of a \emph{single} phase at the core of the system regardless of system size.

To compare the core and local compressibilities we approximate the size of the core region using the Thomas-Fermi radius.  In the Bose-Hubbard model a Thomas-Fermi-type approximation can be used to approximate the density variation with distance near the trap center:
\begin{eqnarray}
n^{\text{TF}}\left ( R_{i,0} \right )=\left ( \mu - \gamma R_{i,0}^{2} \right )/U.
\end{eqnarray}
We focus on parameter regimes which have a density slightly larger than 1 near the center of the trap, e.g., a core superfluid surrounded by a Mott insulator shell with density 1.  The core region is then approximated by a specific radius $R_{\text{c}}$ which encloses sites with particle density larger than 1.  Particles are defined to be in the Thomas-Fermi core of the sample if they sit on sites with $\vert {\bf R}_{i,0} \vert \lesssim R_{\text{c}}$.    We can use the Thomas-Fermi approximation to estimate $R_{\text{c}}$ using $n^{\text{TF}}\left ( R^{\text{TF}}_{c} \right )=1$.  This yields $R^{\text{TF}}_{c}= \sqrt{(U-\mu)/\gamma}$.  We expect $R^{\text{TF}}_{c}\approx R_{\text{c}}$ when a core superfluid is surrounded by a Mott insulator in a trap with a large number of core particles.  We stress that the core compressibility, Eq.~(\ref{corek}), implicitly defines the core region in terms of double occupancies but $R^{\text{TF}}_{c}$ is an approximation we use to compare the local and core compressibilities.

The local compressibility can be combined with our definition of the core of the sample.  By summing the local compressibility over sites only within the core of the sample we construct a compressibility measure of a single phase within the trap:
\begin{equation}
 \kappa_{L}\left ( R_{\text{c}} \right ) =\frac{1}{N_{R_{\text{c}}}}\sum_{ \left \{ i |  R_{i,0} \leq R_{\text{c}}\right \}}\kappa_{i}
 \label{localcompressrc}
\end{equation}
where $ N_{R_{\text{c}}}$ is the number of sites inside the sphere with radius $R_{\text{c}}$.  $\kappa_{L}$ measures the compressibility near the trap center.

We can compare Eq.~(\ref{localcompressrc}) with the core compressibility if we include an appropriate scaling factor.  The core compressibility is defined to be a bulk quantity, summed over all sites and normalized by $N_{s}$ in Eq.~(\ref{corek}).  But the overall $N_{s}^{-1}$ factor in Eq.~(\ref{corek}) is not unique.  To compare $\kappa_{c}^{d}$ with $\kappa_{L}$ we normalize the core compressibility by the approximate number of sites in the core, $ N_{R_{\text{c}}}$, instead of $N_{s}$ :
\begin{equation}
 \kappa_{c}^{d}\left( R_{c} \right)\equiv \frac{1}{2N_{R_{c}}}\frac{\partial }{\partial \mu }
\left \langle \sum_{i}^{N_s} n_{i}\left ( n_{i}-1 \right )\right \rangle.
\label{coreklocal}
\end{equation}
Here the sum is kept over all sites, $N_s$. Because in a trapped system the double occupancy excludes the edge sites.  The sum will then effectively run over $\approx N_{R_{c}}$ sites.  

 \begin{figure}
  \centering
  \includegraphics[width=0.5\textwidth]{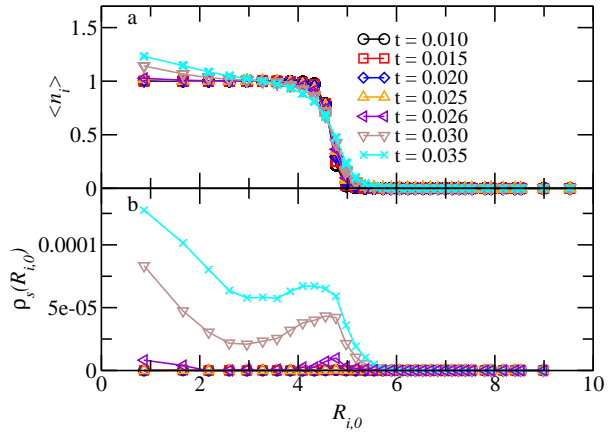}
  \caption{(Color online) Quantum Monte Carlo results showing the local density (a) and superfluid stiffness (b) as a function of the distance from the center, $R_{i,0}$, for different hoppings and $L=12$.  The solid lines are guides to the eye.  Parameter values ($\mu/U =0.76$, $T/U=0.025$, and $\gamma/U=0.035$) are chosen to host a Mott insulator in the core for low hopping but a superfluid in the core for larger hopping.  Note that for $t/U\gtrsim 0.03$ the system also shows a finite local stiffness for all $R_{i,0}$ due to finite size effects.  At these temperatures and system sizes the edge superfluid couples to the core superfluid.  Double occupancy based measures exclude these edge effects. }
  \label{TrappedvsHopping_rhoandn}
\end{figure}

We now use QMC to compare a quantity we interpret mathematically as the local compressibility of the sample core, Eq.~(\ref{localcompressrc}), and a quantity that we propose can be used to effectively measure the core compressibility in experiments, Eq.~(\ref{coreklocal}).   We choose parameters so that the core of the trapped system moves along a line equivalent to ``b'' in Fig.~\ref{phasediagram}.  Fig.~\ref{TrappedvsHopping_rhoandn} plots the local stiffness and local density for several $t/U$ to show that for low $t/U$ we have a Mott insulator in the core but for larger $t/U$ we have a finite superfluid density in the core.  When $t/U$ crosses the phase boundary the density at the trap center becomes larger than 1 and the stiffness at the center becomes larger than zero.

\begin{figure}[t]
  \centering
  \includegraphics[width=0.5\textwidth]{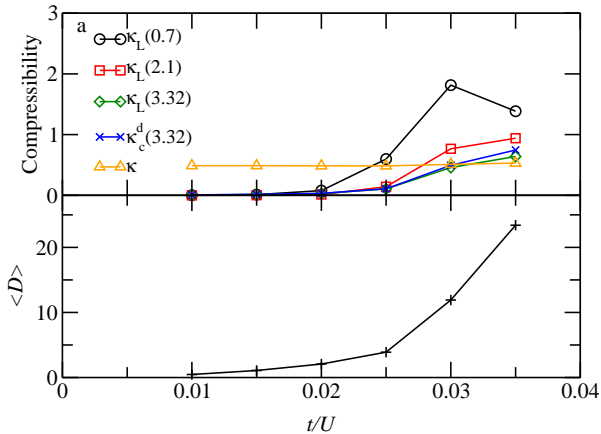}
  \caption{(Color online) Local, $\kappa_{L}\left ( R_{\text{c}} \right )$, total, $\kappa$, and core, $\kappa_{c}^{d}\left( R_{c} \right)$ compressibilities (a) and double occupancy (b) plotted as a function of hopping for the same parameters as Fig.~\ref{TrappedvsHopping_rhoandn}.  The solid lines are guides to the eye.  Here we see that the total compressibility (triangles) remains non-zero and nearly flat across the transition because the system edges remain compressible even when low hoppings yield a Mott insulator in the core. The local and core compressibilities show better agreement near the transition when the same definition for the core region is used, $R^{\text{TF}}_{c}=3.32$.  Triple occupancies lead to deviations between  $\kappa_{L}\left ( 3.32 \right )$ and $\kappa_{c}^{d}\left( 3.32 \right)$ for $t/U\approx0.035$.  }
  \label{TrappedvsHopping_kandd}
\end{figure}

Fig.~\ref{TrappedvsHopping_kandd} compares Eq.~(\ref{localcompressrc}) with Eq.~(\ref{coreklocal}) for the same parameters as Fig.~\ref{TrappedvsHopping_rhoandn}. For the core compressibility we plot $\kappa_{c}^{d}\left( R^{\text{TF}}_{c} \right)$.  We find $R^{\text{TF}}_{c}\approx3.32$ lattice spacings for these parameters.  Note that a non-zero core compressibility signals the onset of superfluidity at the trap center.  But because the edges are always compressible we see very little structure in the total compressibility as the core of the system crosses the phase boundary.  $\kappa_{L}\left ( R_{\text{c}} \right )$ is also plotted for different core radii.  We find $\kappa_{L}\left ( R_{\text{c}} \right )\approx \kappa_{c}^{d}\left( R^{\text{TF}}_{c} \right)$ when $ R_{\text{c}}\approx R^{\text{TF}}_{c}$.  Thus the core compressibility offers a quantitatively accurate measure of local compressibility provided we scale the definition of core compressibility to include the same number of sites in the core.  Here we have chosen $N_{R^{\text{TF}}_{c}}$ because we expect $N_{R^{\text{TF}}_{c}}$ to scale accurately with trapping.   

\begin{figure}
  \centering
  \includegraphics[width=0.5\textwidth]{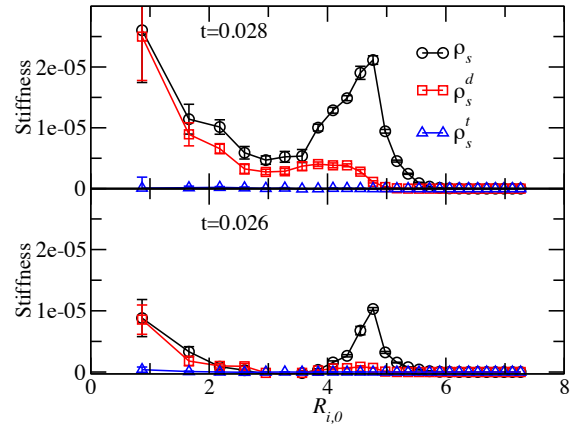}
  \caption{(Color online) Quantum Monte Carlo results plotting the local stiffness (circles) as a function of the distance from the trap center for two different hoppings and $L=16$.  The parameters are otherwise chosen to match Fig.~\ref{TrappedvsHopping_rhoandn}.  The solid lines are guides to the eye.  The top panel shows that strong confinement can enforces a coupling between the core and edge superfluids to yield a non-zero local stiffness throughout the trap.  The bottom panel shows a region of Mott insulator where the stiffness goes to zero near $R_{i,0}=3$. The stiffness arising from doubly occupied sites (squares) and triply occupied sites (triangles) are shown for comparison.  The stiffness near the trap center matches the stiffness from the doubly occupied sites.  The doubly occupied stiffness goes to zero at the trap edges. }
  \label{doublestiffness}
\end{figure}

The core and local superfluid stiffnesses also match near the system center.  Fig.~\ref{doublestiffness} compares the local stiffness to the local stiffness projected onto doubly occupied sites.  The local stiffness in the core matches the stiffness due to doubly occupied sites.  We conclude that the superfluidity in the core is due to coherence among doubly occupied sites.  This implies that doubly occupied sites can also be used to observe core superfluidity.

\begin{figure}[t!]
  \centering
  \includegraphics[width=0.49\textwidth]{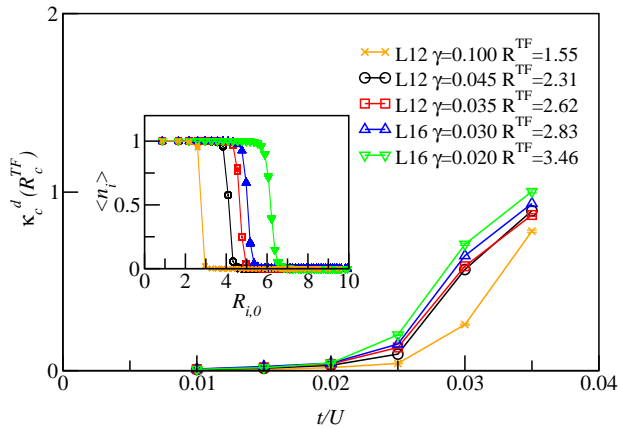}
  \caption{(Color online) Rescaled core compressibility, Eq.~(\ref{coreklocal}), plotted for different trap strengths as a function of hopping. The inset plots the radial dependence of the local density for each system at $t=0.015$.  The other parameter values are $\mu =0.76$ and $T=0.025$.   The data collapse for trap strengths that allow a large core superfluid region, $\gamma \lesssim0.045$.}
  \label{CoreK_differenttrapstrenghts}
\end{figure}

By scaling Eq.~(\ref{coreklocal}) with trapping (and thus increasing particle number) we can extract a bulk measure of core compressibility.  For large enough system sizes an appropriately scaled quantity should show little variation with particle number.  Fig.~\ref{CoreK_differenttrapstrenghts} plots Eq.~(\ref{coreklocal}) for several different values of the trapping strength $\gamma$.  The inset shows that for $\gamma=0.1$ there are very few particles in the core, so the deviation from scaling is the worst in this case. But for smaller values of $\gamma$ a sufficiently large number of particles reside in the core.  The main panel in Fig.~\ref{CoreK_differenttrapstrenghts} indicates that Eq.~(\ref{coreklocal}) shows data collapse provided enough particles reside in the core.  Data collapse for low $\gamma$ indicates that here we can extract the core compressibility in the thermodynamic limit.

In this section we have shown that the core compressibility self-selects sites in the core of the sample.  A local compressibility measure relies on the selection of a finite volume in which to measure the compressibility.  The core compressibility relies on double occupancies to self-select a volume within the center of the trap.  By rescaling a pre-factor used in the local compressibility we showed agreement between the core and local compressibility over a wide range of trap strengths.  We have also shown agreement between the core stiffness and the local stiffness.  

\section{Measuring Core Compressibility}
\label{measuring}

The core compressibility can be extracted from optical lattice experiments to observe phase transitions \cite{jordens:2008,scarola:2009}.  Measurements of the core compressibility rely on variations of the double occupancy with particle number.  The core compressibility ratio $\kappa_{c}^{d}/\kappa$ is given by:
\begin{equation}
\frac{\kappa_{c}^{d}}{\kappa}=\frac{\partial \left \langle D\right \rangle}{\partial \mu}\left (\frac{\partial \left \langle N\right \rangle}{\partial \mu}\right)^{-1}
=\frac{\partial \left \langle D\right \rangle}{\partial \langle N\rangle} .
\label{ccr}
\end{equation}
This ratio is a dimensionless quantity defined entirely in terms of optical lattice observables because the total number of particles, $\langle N \rangle=N_{s}\langle n\rangle$, and the total double occupancy, $D$, are both accessible from time of flight measurements \cite{rom:2004,stoferle:2006}.  Fortunately, edge effects cause $\kappa$ to be a smooth non-zero function in trapped systems even across phase transitions (See, e.g., Fig.~\ref{TrappedvsHopping_kandd}, Ref.~\onlinecite{schneider:2008}, and Ref.~\onlinecite{scarola:2009}).  Thus observation of the slope of the total double occupancy with respect to particle number reveals the core compressibility, up to an overall factor that is nearly constant.

\begin{figure}[t!]
  \centering
  \includegraphics[width=0.5\textwidth]{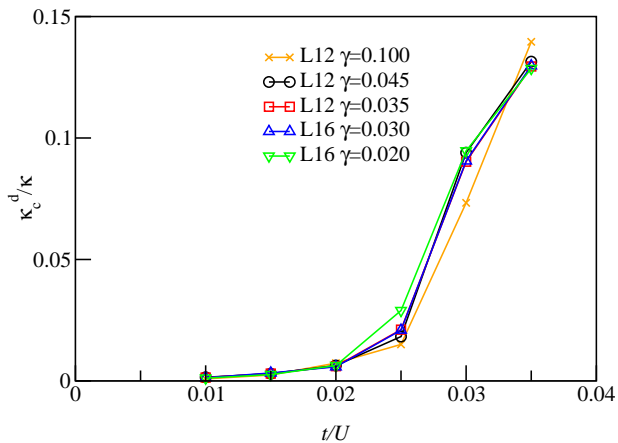}
  \caption{(Color online) The core compressibility ratio plotted for the same parameters as Fig.~\ref{CoreK_differenttrapstrenghts}.  The compressibility of the edges leaves $\kappa$ nearly constant as the core leaves the Mott insulator with increasing hopping.}
\label{Ratiokcoverk}
\end{figure}

Fig.~\ref{Ratiokcoverk} demonstrates that the core compressibility ratio reveals the Mott insulator-superfluid transition.  Here the transition manifests in a quantity that does not rely on microscopic definitions of the core.  The core compressibility ratio is instead defined entirely in terms of bulk values, double occupancy and particle number, accessible from time of flight measurements.

\section{Summary}

Measurements of double occupancy reveal properties at the core of optical lattice experiments even though the experiments come with significant spatial inhomogeneity due to trapping.    Trapping can mix states within one system.  The double occupancy offers a bulk observable that can be used to extract information regarding the core of the sample. We defined core quantities, the core compressibility and the core stiffness, in terms of double occupancy.

We focused our study on a validation of core compressibility as a useful tool in extracting compressibility from experiments on trapped systems.  A measure of core compressibility,  Eq.~(\ref{ccr}),  can be defined entirely in terms of optical lattice observables and therefore offers a powerful experimental method for mapping out phase diagrams even in the presence of trapping.  We first studied the boson core compressibility in a uniform system.   We used MFT and QMC to argue that the core compressibility defined in terms of double occupancy excludes low densities while tracking the total compressibility.  The relationship between double occupancy-based core and total compressibility holds if: i) the number of particles per site is just above unity and ii) Thermal and quantum fluctuations do not allow significant triple occupancy.

We have also studied the core compressibility in trapped systems.  We showed that the core compressibility tracks the change of state within the core of the system while excluding edge effects (unlike the total compressibility which includes edge effects).  We further defined a site normalization that allowed a comparison between local compressibility and the core compressibility.  The core and local compressibility were shown to be essentially the same when compared over the same number of sites.  We conclude that the core compressibility is thus a powerful but simple observational tool that can be used to observe the same critical properties as local compressibility.

The boson core compressibility has been tested on an experimentally relevant model, the trapped Bose-Hubbard model.  The boson core compressibility can be applied more generally to study transitions between a wide variety of incompressible and compressible phases.  One of the most pressing \cite{pasienski:2010} examples is the Mott insulator to Bose-Glass transition \cite{fisher:1989}.  
 
We acknowledge support from the Jeffress Memorial Trust (J-992), AFOSR (FA9550-11-1-0313), and DARPA-YFA (N66001-11-1-4122).
Some of the calculations were performed at the Lonestar cluster in the Texas Advanced Computing Center at the University of Texas at Austin.

\end{document}